\documentclass[aps,prb,groupedaddress,showpacs,twocolumn]{revtex4}
\usepackage{graphicx}
\usepackage{amsmath}
\usepackage{amssymb}
\usepackage{bbm}
\usepackage{xspace}
\usepackage{color}
\usepackage{footnote}

\newcommand{\matx}{\mat{x}}

\definecolor{orange}{RGB}{252,77,6}
\definecolor{brown}{RGB}{200,127,50}
\definecolor{green1}{RGB}{00,100,00}
\definecolor{green2}{RGB}{00,150,00}
\definecolor{green3}{RGB}{00,200,00}
\definecolor{green4}{RGB}{00,250,00}

\newcommand{\fig}[1]{Fig.\thinspace{}\ref{#1}}

\newcommand{\eq}[1]{Eq.\thinspace{}(\ref{#1})}

\newcommand{\se}{Sec.\@\xspace}

\newcommand{\app}{App.\@\xspace}

\newcommand{\etal}[0]{\textit{et al.}}
\newcommand{\tcite}[1]{Ref.~\onlinecite{#1}}

\def\ket#1{\mathinner{|{#1}\rangle}}

\newcommand{\nag}{{\phantom{\dag}}}

\newcommand{\mat}[1]{\mathsf{#1}}

\begin{document}


\title{Steady state and quench dependent relaxation of a quantum dot coupled to one-dimensional leads}


\author{Martin Nuss}
\email[]{martin.nuss@student.tugraz.at}
\affiliation{Institute of Theoretical and Computational Physics, Graz University of Technology, 8010 Graz, Austria}
\author{Martin Ganahl}
\affiliation{Institute of Theoretical and Computational Physics, Graz University of Technology, 8010 Graz, Austria}
\author{Hans Gerd Evertz}
\affiliation{Institute of Theoretical and Computational Physics, Graz University of Technology, 8010 Graz, Austria}
\author{Enrico Arrigoni}
\affiliation{Institute of Theoretical and Computational Physics, Graz University of Technology, 8010 Graz, Austria}
\author{Wolfgang von der Linden}
\affiliation{Institute of Theoretical and Computational Physics, Graz University of Technology, 8010 Graz, Austria}


\date{\today}

\begin{abstract}
We study the time evolution and steady state of the charge current in a Single Impurity Anderson Model, using Matrix Product States techniques. A non equilibrium situation is imposed by applying a bias voltage across one-dimensional tight binding leads. Focusing on particle-hole symmetry, we extract current-voltage characteristics from universal low bias up to high bias regimes, where band effects start to play a dominant role. We discuss three quenches, which after strongly quench dependent transients yield the same steady state current. Among these quenches we identify those favorable for extracting steady state observables. The period of short time oscillations is shown to compare well to real-time renormalization group results for a simpler model of spinless fermions. We find indications that many body effects play an important role at high-bias-voltage and finite bandwidth of the metallic leads. The growth of entanglement entropy after a certain time-scale $\propto \Delta^{-1}$ is the major limiting 
factor for calculating the time evolution. We show that the magnitude of the steady state current positively correlates with entanglement entropy. The role of high energy states for the steady state current is explored by considering a damping term in the time evolution. 
\end{abstract}


\maketitle

\section{Introduction}\label{sec:introduction}
Over the past decade, experimental control over quantum systems has increased considerably. Possible realizations reach from model Hamiltonians~\cite{Trotzky18012008, schneider_fermionic_2012} using ultra cold atoms in optical lattices to experimental setups of nanoscopic devices like molecular junctions, quantum wires or quantum dots.~\cite{goldhaber_from_1998, PhysRevB.85.201301} Many of these systems show remarkable properties, often due to reduced effective dimensionality and many body interactions. A prominent example is the Kondo effect,~\cite{hewson_kondo_1997} which plays an essential role in transport across quantum dots. A theoretical understanding of transport in out of equilibrium conditions is highly interesting for applications in nano- and molecular- electronics and even in biological systems.

Electron-electron interactions render the theoretical description of non equilibrium dynamics one of the most challenging problems in today's condensed matter physics.~\cite{PhysRevB.48.8487} However, with the advent of efficient numerical techniques to simulate one-dimensional (1d) quantum systems,~\cite{PhysRevB.48.10345,daley_tdmrg, PhysRevLett.93.040502, white_feiguin_tdmrg,Schollwoeck201196,PhysRevB.70.121302} many physical problems are well within grasp of theoretical physicists. Even non equilibrium setups in regimes where the potential bias is large with respect to the energy scales of the unperturbed systems are now feasible to study.~\cite{PhysRevLett.101.140601,PhysRevB.79.235336,PhysRevLett.107.206801}

In this work we obtain the steady state charge current of a single interacting quantum dot under voltage bias, modeled by a single-impurity Anderson model (SIAM).~\cite{anderson_localized_1961} This model is commonly discussed in the wide-band limit~\cite{footnote1} approximation, tailored towards a universal, low-bias transport description. Here, we extend the discussion to the case of a finite (semi-circular) conduction band in the leads, which has not been explored specifically. A particular realization could consist of two one-dimensional leads like nano-wires~\cite{Dekker1997,Hasegawa2001367,PhysRevLett.108.176802,PhysRevLett.109.156804} and a junction between them comprised of a magnetic impurity i.e. the quantum dot. We use generic one-dimensional tight binding leads with finite electronic bandwidth which mimic the electronic properties of for example carbon nano-tubes~\cite{APL.60.2204}. In such a device the electronic density of states (DOS) of the leads would have a bandwidth on the order of $15\,
eV$~\cite{APL.60.2204,barford_2005} and effects arising from their specific structure are to be expected when using corresponding bias voltages which are larger than those typically applied in current experiments with nanoscopic devices.

The steady state is obtained by combining Density Matrix Renormalization Group (DMRG)~\cite{PhysRevB.48.10345,Schollwoeck201196} and Time Evolving Block Decimation (TEBD)~\cite{PhysRevLett.93.040502,Schollwoeck201196} techniques, to perform real time evolution of the system after several different quenches. This technique is known to yield reliable results for a wide parameter range of one dimensional models~\cite{PhysRevLett.101.140601,PhysRevB.79.235336,PhysRevB.70.121302,PhysRevB.73.195304,PhysRevB.82.205110,PhysRevLett.106.220601,PhysRevB.84.174438,MeisnerEPJB2009,JPSJ.77.084704,JPSJ.79.093710,PhysRevLett.88.256403,PhysRevLett.91.049701,PhysRevLett.91.049702,PhysRevLett.101.236801} and to agree with analytical data.~\cite{PhysRevLett.101.140601}

We focus on the particle-hole symmetric point which shows the most pronounced many body effects.~\cite{PhysRevB.85.235107} The bias voltage for most of our data is much larger than the equilibrium Kondo temperature (see \se~\ref{ssec:kondo}), so that Kondo correlations should not influence the steady state current. We show that the same steady state current is reached independent of the type of quench used and identify quenches which are superior to others when it comes to extracting steady state data. We investigate quench induced oscillations in the transients and compare to real-time Renormalization Group results. We have performed a careful convergence study in all auxiliary numerical and system parameters and found the major limitation to be the truncation of the many-body state space in each iteration. The method is well suited for reaching relevant time scales to study the steady state current. We find that our approach is capable of yielding unbiased results valid in the thermodynamic limit. Data 
presented in this work reproduces analytical results in the non interacting system. In the low bias region our results for the current-voltage characteristics agree with previous data (Heidrich-Meisner \etal$\;$\tcite{PhysRevB.79.235336}). We are able to extend earlier results~\cite{PhysRevB.81.035108,PhysRevB.68.155310,PhysRevB.79.235336,jakobs_nonequilibrium_2010,PhysRevB.86.245119,Dirks2012b,PhysRevB.77.195316} to a wider parameter regime and discuss the interplay of finite lead bandwidth and electronic correlations. We find evidence for pronounced many body effects at high bias voltages in interplay with finite electronic bandwidths of the leads.~\cite{PhysRevB.81.193401} Finally we discuss the role of high energy states for the steady state current in low- and high bias voltage regimes.

The text is organized as follows: In \se~\ref{sec:model} we introduce our model and describe in detail the different quenches to be performed. We present data for the transient response in \se~\ref{sec:transient}. Results for the steady state current are presented in \se~\ref{sec:results}  where we also outline how to extract steady state data from time evolved quantities. We analyze time scales of individual parameter regimes in \se~\ref{ssec:timePerformance}. The role of high energy states in different bias regimes is discussed in \se~\ref{ssec:resultsDamped}. A detailed convergence analysis is presented in \app~\ref{app:preliminaryConsiderations}.

\section{Setup}\label{sec:model}
In this section we define our notation for the SIAM which we use to model a quantum dot. We are interested in electron transport across the quantum dot after each of the several quenches to be described in detail in the following. We explain how we calculate the ground state using DMRG and the real time evolution using TEBD.

\begin{figure}
  \centering
  \includegraphics[width=0.49\textwidth]{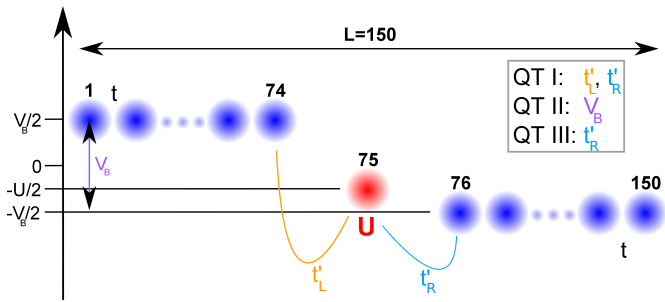}
  \caption{(Color online) Illustration of the three quenches performed for the SIAM: i) QT I: quenching of both quantum dot-reservoir tunnelings $t'_{L}$ and $t'_{R}$, ii) QT II: quenching the bias voltage $V_B$ and iii) QT III: quenching the dot-lead tunneling $t'_{R}$ to one lead.}
  \label{fig:QD_TEBD_QT}
\end{figure}

\subsection{Single-impurity Anderson model}\label{ssec:modelSIAM}
We consider a model for a quantum dot including charge as well as spin fluctuations described by the SIAM, consisting of an
interacting site connected to a bath of non interacting electrons. We choose a setup where the quantum dot is located in the middle of a one-dimensional chain of tight-binding electrons. The dynamics are governed by
\begin{subequations}
\begin{align}
\hat{\mathcal{H}} &= \hat{\mathcal{H}}_{\text{dot}} + \hat{\mathcal{H}}_{\text{res}} + \hat{\mathcal{H}}_{\text{coup}} \label{eq:HSIAM}\\
\hat{\mathcal{H}}_{\text{dot}} &= -\frac{U}{2} \, \sum\limits_{\sigma}  \, f_{\sigma}^\dagger \, f_{\sigma}^\nag + U \, \hat{n}^{f}_{\uparrow} \, \hat{n}^{f}_{\downarrow}  \label{eq:Hf}\\
\hat{\mathcal{H}}_{\text{res}} &= \sum\limits_{\alpha,\sigma}\,\left( \epsilon_{\alpha} \, \sum\limits_{i}  \,c_{i\alpha\sigma}^{\dagger} \, c_{i\alpha\sigma}^{\nag} - t \, \sum\limits_{\left\langle i,\,j \right\rangle}  \, c_{i\alpha\sigma}^{\dagger} \, c_{j\alpha\sigma}^{\nag}\right) \label{eq:Hres}\\
\hat{\mathcal{H}}_{\text{coup}} &= -\sum\limits_{\alpha} t_{\alpha}'\sum\limits_{\sigma}\left(c_{0\alpha\sigma}^{\dagger} \, f_{\sigma}^\nag + f_{\sigma}^\dagger \, c_{0\alpha\sigma}^{\nag} \right)\;\mbox{,}
 \label{eq:Hcpl}
\end{align}
\end{subequations}
(see \fig{fig:QD_TEBD_QT}) where $U$ parametrizes the on-site interaction strength on the quantum dot, $t'_{\alpha},\alpha\in\{{\rm L,R}\}$ is the coupling strength between the quantum dot and the left and right lead. Lead $\alpha$ is characterized by intra lead hopping $t$ and on-site potential $\epsilon_{\alpha}$. Particle-hole symmetry is enforced for all chosen parameters. When needed, the on-site energy of the quantum dot will be denoted by $\epsilon_f$.

We choose $t=1$ and symmetric couplings $t'_{L}=t'_{R}=0.3162\,t$ (\eq{eq:Hcpl}) for all simulations, yielding a bandwidth of $D=4\,t$ of the leads and an equilibrium Anderson width~\cite{hewson_kondo_1997} of $\Delta \equiv \pi\,t'^2_{\alpha} \,\rho_{\text{reservoir}}(0) = \frac{t'^2_{\alpha}}{t} \approx 0.1\,t$. We choose $t=1$ and symmetric couplings $t'_{L}=t'_{R}=0.3162\,t$ (\eq{eq:Hcpl}) for all simulations. This yields a bandwidth of $D=4\,t$ of the leads and an equilibrium Anderson width~\cite{hewson_kondo_1997} of $\Delta \equiv \pi\,t'^2_{\alpha} \,\rho_{\text{reservoir}}(\mu) = \frac{t'^2_{\alpha}}{t} \approx 0.1\,t$, where the reservoir DOS at the chemical potential is denoted by $\rho_{\text{reservoir}}(\mu)$. We will display all energies in units of $\Delta$ ($\hbar, k_B \mbox{ and } e = 1$). We restrict ourselves to the zero temperature case. Real time will be denoted by $\tau$. In \app~\ref{app:preliminaryConsiderations} we show that within the simulation time $\tau$ accessible, the 
finiteness of the leads does not affect our results.

\subsection{Quench preparation}\label{ssec:modelQUENCH} 
We are interested in the steady state current~\cite{footnote2} of \eq{eq:HSIAM} under a finite bias voltage $V_B$.~\cite{0034-4885-70-2-R02,JPSJ.65.30} Our strategy to obtain the steady state is by quenching the Hamiltonian parameters $\matx_0=\{U, t, t'_{\alpha}, \epsilon_\alpha\}$ at $\tau=0$ from some initial to their final values $\hat{\mathcal{H}}(\matx_0)\rightarrow \hat{\mathcal{H}}(\matx)$ and evolve an initial state $\ket{\Psi_0}$ with $\hat{\mathcal{H}}(\matx)$. $\ket{\Psi_0}$ is chosen to be the ground state of the initial Hamiltonian $\hat{\mathcal{H}}(\matx_0)$ in the canonical ensemble at half filling with total spin projection $S^z=0$.

It has been shown that the steady state is independent of the quench rate.~\cite{JPSJ.79.093710,PhysRevLett.88.256403} We apply all quenches at an instant of time i.e. without a ramp. It could however be interesting to study the entanglement growth as a function of the quench ramp.

We consider three different quench types (see \fig{fig:QD_TEBD_QT}) which will be explained in detail below. Unless stated otherwise, we choose a system of $L=150$ sites with the quantum dot located at site $75$. To drive the system out of equilibrium, a bias voltage $V_B$ is applied by setting the respective on-site energies of the leads in an anti-symmetric fashion to $\epsilon_L =-\epsilon_R = \frac{V_B}{2}$. For all quenches, the final parameters are $\matx=\{U,t=1,t'_{\alpha}=0.3162\,t,\epsilon_L=\frac{V_B}{2}, \epsilon_R=-\frac{V_B}{2}\}$, with variable $U$. The initial setup is quench type (QT)-dependent (see \fig{fig:QD_TEBD_QT}):

\subsubsection{QT I: Hybridization quench to both leads $t'_{\alpha}=0\rightarrow 0.3162\,t$}\label{sssec:QTI} 
For $\tau<0$ we take $\matx_0=\{U,t,t'_{\alpha}=0,\epsilon_{\alpha}=\pm V_B/2\}$, i.e. no quantum dot-to-leads coupling, but the bias voltage is already applied. We prepare the ground state of $\hat{\mathcal{H}}(\matx_0)$ at half-filling in the left and right lead and a single up-electron on the quantum dot. At $\tau = 0$ the tunneling $t_{\alpha}'$ is quenched to its finite value. Note that due to the splitting into three disconnected parts ($t'_{\alpha}=0$), $S^z$ is not zero on the quantum dot and on the right lead initially.

\subsubsection{QT II: Quenching the bias voltage $\epsilon_{\alpha}=0\rightarrow \pm V_B/2$}\label{sssec:QTII} 
At $\tau<0$, $\matx_0=\{U,t,t'_{\alpha}=0.3261\,t,\epsilon_{\alpha}=0\}$. The system
is prepared in the ground state $\ket{\Psi_0}$ at half filling with
overall $S^z=0$ zero. At $\tau= 0$ the bias voltage is quenched to
its desired value. As compared to QT I, this setup has the advantage
that no subsystems with finite values of $S^z$ exist in the
ground state. Furthermore, correlations between the three regions are already present in the ground state. Note that the initial state is much more complicated than for QT I. This type of quench has also been used by the authors of \tcite{PhysRevB.82.205110}.

\subsubsection{QT III: Quenching the hybridization $t'_{R}=0\rightarrow 0.3162\,t$ to the right lead}\label{sssec:QTIII}
The initial parameters are chosen $\matx_0=\{U,t,t'_{L}=0.3261\,t,t'_{R}=0,\epsilon_{\alpha}=\pm V_B/2\}$, and the system is again solved for the ground state $\ket{\Psi_0}$ at half filling. At $\tau=0$, we quench $t'_R=0\rightarrow 0.3162\,t$ and evolve $\ket{\Psi_0}$ with the quenched Hamiltonian.

\subsection{Methods}\label{ssec:methods}
To prepare the system in the ground state of the initial Hamiltonian, we employ the DMRG~\cite{PhysRevB.48.10345} algorithm in its two and single site formulation. Our implementation exploits conservation of spin projection ($S^z$) and charge ($N$), which is crucial for obtaining high precision data. Time evolution is done using the TEBD~\cite{PhysRevLett.93.040502} algorithm, within a second order Suzuki-Trotter decomposition of the propagator
\begin{align*}
  e^{-i\hat{\mathcal{H}} T} = \left(e^{-i\hat{\mathcal{H}}\delta\tau}\right)^{\frac{T}{\delta\tau}} = \left(e^{\frac{{\delta\tau}}{2}\hat{\mathcal{H}}_o}e^{{\delta\tau}\hat{\mathcal{H}}_e}e^{\frac{{\delta\tau}}{2}\hat{\mathcal{H}}_o}\right)^{N_{\tau}} + \mathcal{O}\left(\delta\tau^3\right)\;\mbox{,}
\end{align*}
where $N_{\tau} = \frac{T}{\delta\tau}$ is the number of time slices, $T$ is the total simulation time and $\delta \tau$ the length of a single time step. The operators $\hat{\mathcal{H}}_e$ and $\hat{\mathcal{H}}_o$ act on even and odd bonds of the bipartite lattice respectively. Unless stated otherwise we use a TEBD matrix dimension of $\chi_{\text{TEBD}}=2000$ and a Trotter time step of $\delta\tau =0.05\,t^{-1}$. For additional details including studies of convergence in system size $L$ and all auxiliary numerical parameters we refer the reader to \app~\ref{app:preliminaryConsiderations}.

The calculations carried out in this work set very high computational demands ($\approx$ one million CPU hours) and were only possible due to a parallel code~\cite{PhysRevLett.108.077206,ZaunerCoMoving,KnapBound2011} which respects quantum number ($N$, $S_z$) conservation.

\section{Transient response}\label{sec:transient}
In this section we present results for the transient current response of the three quenches. We discuss individual bias regimes and identify oscillations in the time evolution of the current which are reminiscent of results for an interacting resonant level model of spinless fermions. We show that QT II leads to much larger initial oscillations than the other two QTs.

\begin{figure*}
\includegraphics[width=0.99\textwidth]{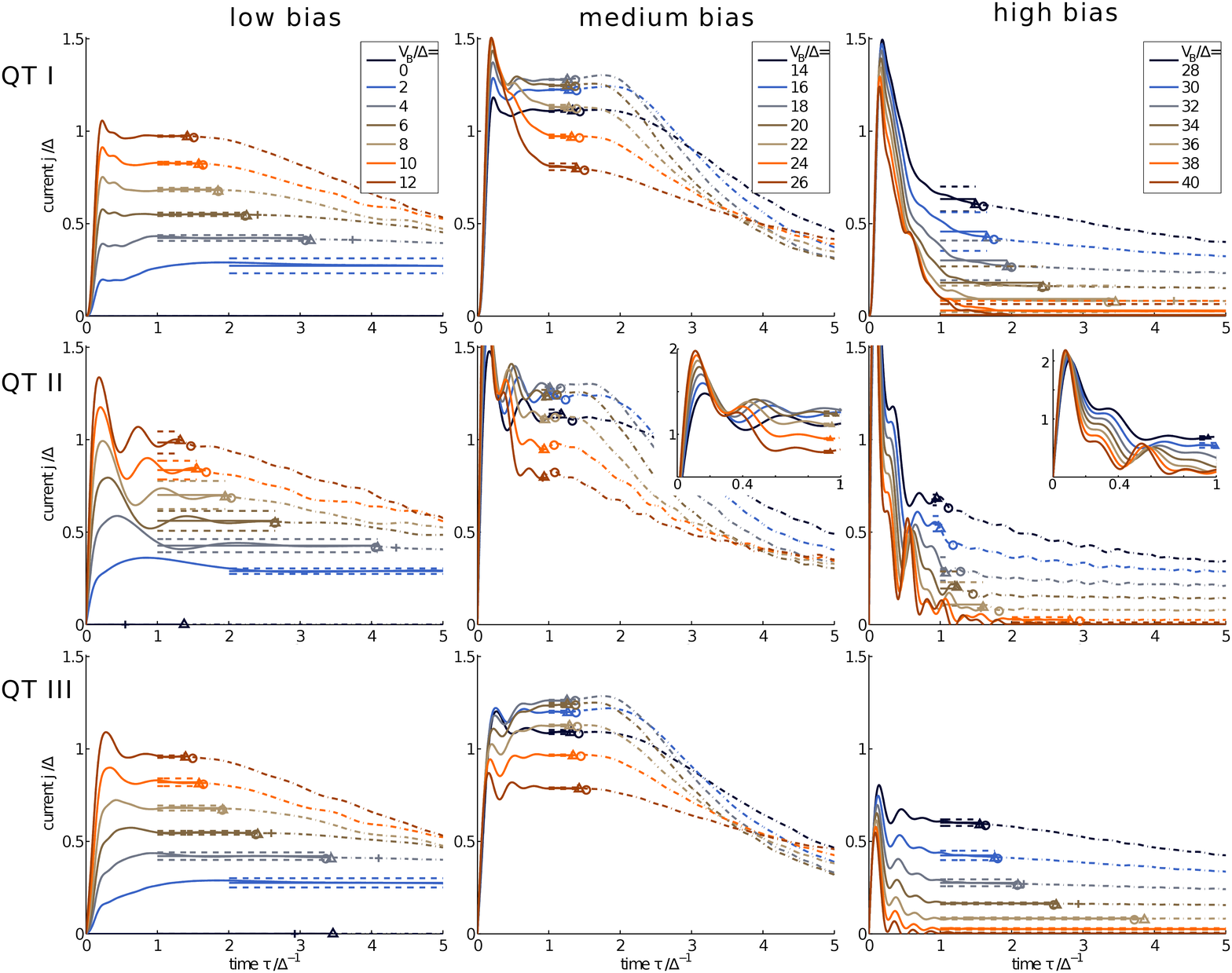}
\caption{(Color online) Time dependence of the current (\eq{eq:currentOperator}) at $U/\Delta=12$ for the three different QTs and for different bias voltages. The curves are plotted as solid lines up to the last  reliable point in the TEBD calculation (see text). Larger times are plotted as dash-dotted lines. Solid horizontal lines are fits to extract the steady state currents. The time domain for these fits  starts at $\tau\approx\Delta^{-1}$ and ends at a point which is identified as the last reliable data point (symbols, see text). Dashed horizontal lines indicate the uncertainty. The insets in the mid row show respective zooms onto short time regions, which are not visible in the main part of the figure.} 
\label{fig:TEBDcurrentOfTime}
\end{figure*}

\subsection{Low, medium and high bias regime}\label{ssec:lowmidhigh}
In our simulations we identify three regimes of bias voltage $V_B$ with qualitatively different behavior. Within each regime, the general features of the time evolution of the current are only moderately dependent on interaction strength. For that reason, we first discuss results for $U/\Delta=12$ only (see \fig{fig:TEBDcurrentOfTime}).

For low bias voltages ($V_B/\Delta \in (0,18)$), a steady state current plateau~\cite{PhysRevB.85.235141,PhysRevB.48.8487} is reached within $\tau\approx\Delta^{-1}$.

In a region of medium bias voltages ($V_B/\Delta\in(18,28)$) we observe a fast increase in current over a timescale of $\tau\approx 0.3\,\Delta^{-1}$ followed by a rather slow decay which, for some model parameters, is too slow to reach a steady state plateau within accessible simulation times (see below).

In the high bias region ($V_B/\Delta\in(28,40)$) the time evolution of the current shows a sharp peak followed by fast decrease of the current into a steady state plateau within $\tau\approx\Delta^{-1}$.

Our data indicates that within a simulation time of $\tau=3\,\Delta^{-1}$ approximately one particle is transferred from the left reservoir to the right one. As discussed in detail in \se~\ref{sec:results}, all three QT eventually approach the same steady state, although in quite different manner. QT II for example leads to the largest transient current spike, which is one reason for the lower accuracy in determining the steady state for this quench. We also find that quenching the hybridization(s) (QT I or III) yields much cleaner steady state plateaus as compared to quenching the bias voltage (QT II), which leads to more pronounced oscillations in these plateaus. 

\begin{figure}
\includegraphics[width=0.49\textwidth]{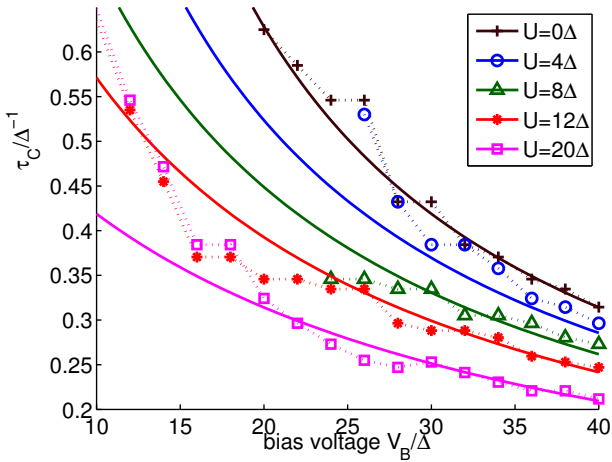}
\caption{(Color online) Period of the sinusoidal oscillations of the current in QT II for various values of interaction strength $U/\Delta=0,4,8,12\mbox{ and }20$ (symbols). Solid lines indicate the predicted form for the interacting resonant level model.~\cite{PhysRevB.83.205103}}
\label{fig:currentRelax}
\end{figure}

\subsection{Characteristic oscillations of the current}\label{sssec:currOszi}
The time evolution of the current exhibits oscillations which are more or less pronounced depending on the type of quench. These oscillations become more explicit with increasing interaction strength (not shown). Their period is of the order of $0.5\,\Delta^ {-1}$ for low bias voltages and decreases to about $0.3\,\Delta^ {-1}$ for higher bias voltages, in a range of interaction strengths $U=[0,20]\,\Delta$. These oscillations compare nicely to results from real-time renormalization group for the interacting resonant level model (see \tcite{PhysRevB.83.205103}, equation $107$), which predicts a sinusoidal behavior ($\propto\sin{(\frac{\tau}{\tau_C}})$) with a period of
\begin{align*}
 \tau_C(U,V) = \frac{2}{V_B+U}\;\mbox{.}
\end{align*}
In \fig{fig:currentRelax}, we plot $\tau_C(U,V)$ as a function of interaction strength and find remarkable agreement with rtRG-results at higher bias voltages. The period was extracted from the TEBD time evolution data in three ways: i) by fitting a sine function, ii) by identifying the dominant Fourier amplitudes and iii) by identifying local maxima. These data were combined and their individual uncertainty taken into account. Error bars in \fig{fig:currentRelax} (not shown) would be sharply growing below $V_B=25\,\Delta$. In the lower bias regime our data is not significant for a reliable extraction of the period.

\begin{figure}
\includegraphics[width=0.49\textwidth]{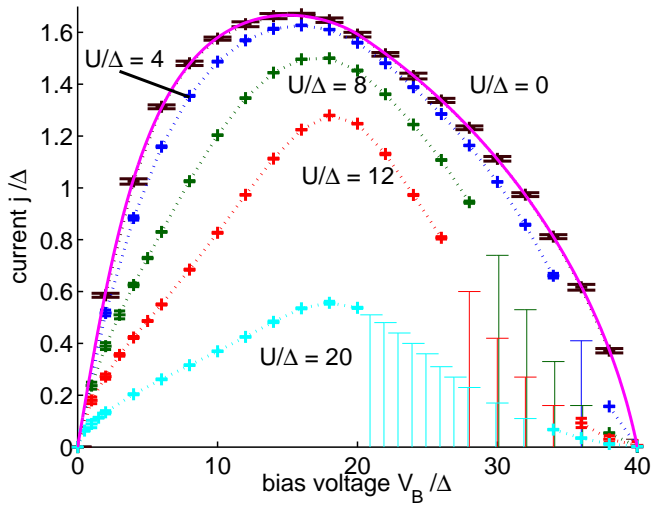}
\caption{(Color online) Current-voltage characteristics of the quantum dot. The steady state currents shown are obtained by a fit of the expectation value of the current operator within the steady state plateau. Regions where only a likely upper bound for the steady state current could be obtained are indicated by pedestals (see text).}
\label{fig:TEBDcurrent}
\end{figure}

\section{Steady state current}\label{sec:results}
In this section we present the current-voltage characteristics of quantum dot. We outline a scheme to extract the steady state current and investigate the dependence on the type of quench used. The current-voltage characteristics in the low bias region is compared to existing data obtained with other methods. Furthermore we present a detailed comparison between an interacting and a non interacting quantum dot for finite as well as infinite lead bandwidth.

\subsection{Extracting the steady state current}\label{ssec:extracting}
We identify the steady state current as the mean value of the time-dependent current taken over a suitable time domain $[\tau_S,\tau_E]$ over which the current shows an almost constant behavior (apart from small oscillations). $\tau_S$ typically depends on the model parameters and was chosen by hand, and $\tau_E$ is taken to be the largest time for which simulations yield reliable results (see \fig{fig:TEBDcurrentOfTime}). Beyond $\tau_E$ the current becomes numerically unreliable, resulting in an artificially decaying current (see \app~\ref{app:preliminaryConsiderations} for discussion). We find that in most of the parameter regions the transients have decayed at $\tau_S\approx\Delta^{-1}$. On the other hand, the end point of the plateau strongly depends on the parameter region under consideration. We define it by two distinct measures. One is the time $\tau^{(1)}_E$ for which the truncated weight $\epsilon$ (see \eq{eq:tw}) reaches a threshold of $\epsilon_c=3\cdot10^{-6}$ at any bond (marked by $+$ in \
fig{fig:TEBDcurrentOfTime}). The second definition ($\tau^{(2)}_E$ marked by $\circ$ in \fig{fig:TEBDcurrentOfTime}) is given by the time for which two different definitions of the current, namely the expectation value of the current operator (\eq{eq:currentOperator}) and  the time derivative of the particle number (\eq{eq:currentDerivative}), deviate by more than $7\cdot10^{-4}$, the latter  being more susceptible to accumulation of errors. Both times are in good agreement with each other and can be combined into an effective simulation time $\tau_E=\text{min}(\tau^{(1)}_E,\tau^{(2)}_E)+\alpha|\tau^{(1)}_E-\tau^{(2)}_E|$ (marked by triangles in \fig{fig:TEBDcurrentOfTime}). We choose a value of $\alpha=0.1$. Results do not depend on this particular choice. It turns out that this procedure is very robust and does also agree with the point at which the TEBD current starts to deviate from the exact time evolution in the non interacting system (see \app~\ref{ssec:ctar}).

The steady state plateaus obtained in this way usually show oscillations and/or small, parameter- and quench-dependent drifts. We quantify the quality of convergence within the plateau region $[\tau_S,\tau_E]$ by the slope of a linear fit to the current. A large slope indicates that it is not possible to reach the steady state within the given simulation time $\tau_E$, i.e. the physical relaxation time is too long or the reached simulation time is too short. This is further discussed in \se~\ref{ssec:timePerformance}. For these parameter values we can only provide a likely upper bound for the steady state current, given by the current at the last reliable simulation time. This is justified because we find the current to always decrease as a function of time (apart from small oscillations). Note that although for some of these parameters the current in some QT's may appear converged but is still considered not converged according to our strict criteria. We consider the current to be converged when the 
relative slope is below a threshold of $\approx 5\cdot 10^{-2}\,\Delta$. Each curve in addition was inspected by hand for convergence. When we consider the steady state current converged, we estimate its error as three times the standard deviation taken over the data points in the fitting interval $[\tau_S,\tau_E]$ (plotted as dashed lines in \fig{fig:TEBDcurrentOfTime} and ~\fig{fig:TEBDcurrent}). This coincides most of the time with the maximal deviation of the time-dependent current from its mean value.

As an important test, we obtained the current-voltage characteristics for the non interacting case and compared it to analytical results~\cite{footnote3} (see \fig{fig:TEBDcurrent}), finding excellent agreement (see also \app~\ref{ssec:ctar}). Another indication for the reliability of the scheme outlined above is that all three types of quenches investigated yield the same steady state current within the uncertainty. We note that this is not a priori clear since quench dependent steady states have been reported in different systems.~\cite{PhysRevLett.105.156802} As noted in \app~\ref{app:preliminaryConsiderations} the position of the plateau is also stable with respect to variations of technical parameters of the simulation. The quality of the steady state plateau, however, depends strongly on the values of interaction and bias voltage and may be obscured by initial oscillations or shortened at the end by the truncated weight breakdown.

The behavior of the spin-current strongly depends on the quench type and it is even identical to zero for QT II. In this respect, the steady state charge current does not depend on the properties of the spin current, since all three quenches yield the same steady state for the charge current. This turns out to be very advantageous since the time scales in the spin sector are much larger than in the charge sector.~\cite{Cohen2012,NussGanahlEtcInPreparation}

From our calculations, we find QT I and QT III to yield more reliable data for the extraction of the steady state current than QT II. Reasons for this behavior are i) the much more pronounced oscillations in the data of QT II which enlarge the statistical uncertainty of steady state values and ii) the much higher transient spike in QT II accompanied by a slightly higher initial entanglement and shorter $\tau_E$. Entanglement growth is in general parameter dependent and converges towards the same value for all quench types.~\cite{NussGanahlEtcInPreparation} In the following, we will present steady state data extracted from QT I and QT III.

\subsection{Current-voltage characteristics}\label{ssec:resultsss}
\begin{figure}
\includegraphics[width=0.49\textwidth]{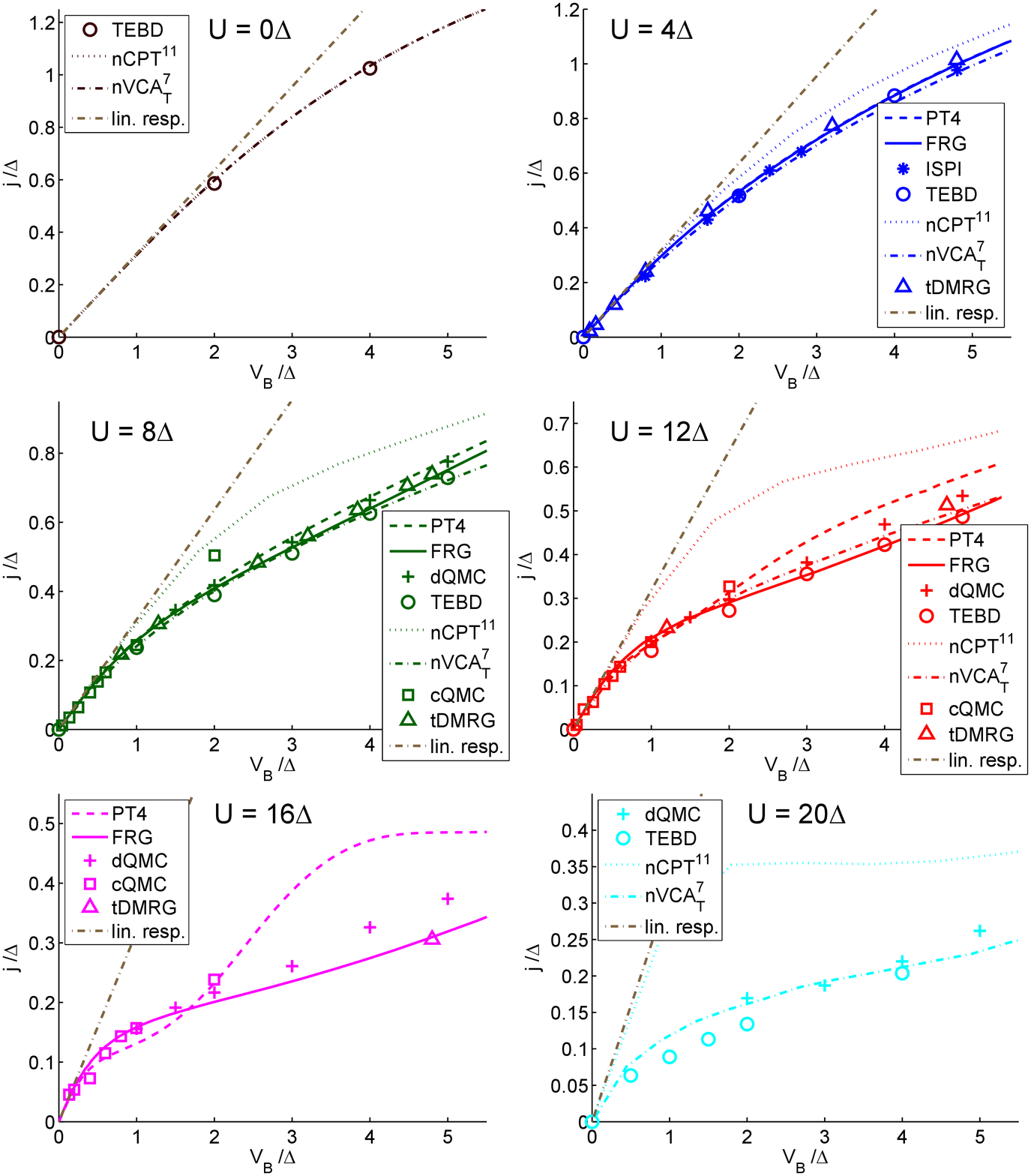}
\caption{(Color online) Comparison of the current-voltage
  characteristics of the SIAM obtained with different methods in the low-bias regime. Some of the methods use a wide-band limit and others a semi-circular reservoir DOS which (for equal $\Delta$) become comparable in the low bias region shown. The methods  are: 1) diagrammatic QMC for $T=0$ in the wide-band limit (dQMC),~\cite{PhysRevB.81.035108} 2) fourth order Keldysh perturbation theory for $T=0$ in the wide-band limit (PT4),~\cite{PhysRevB.68.155310} 3) time-dependent DMRG for $T=0$ using a semi-circular DOS (tDMRG),~\cite{PhysRevB.79.235336} 4) TEBD for $T=0$ using a semi-circular DOS (TEBD, this work), 5) non equilibrium FRG for $T=0$ using a wide-band limit (FRG),~\cite{jakobs_nonequilibrium_2010} 6) non equilibrium Cluster Perturbation Theory for $T=0$ using a semi-circular DOS (nCPT$^{11}$),~\cite{PhysRevB.86.245119} 7)
  non equilibrium Variational Cluster Approach for $T=0$ using a
  semi-circular DOS
  (nCPT$^{7}_{T}$),~\cite{PhysRevB.86.245119} 
8) imaginary time QMC for $T=0.2\,\Delta$ in the wide-band limit
  (cQMC)~\cite{Dirks2012b} 9) iterative summation of real-time
  path integrals for $T=0.2\,\Delta$ in the wide-band limit
  (ISPI)~\cite{PhysRevB.77.195316} and 10) the linear response
  result for the Kondo regime $j_{\text{lin}}=2 G_0 V_B$ (lin. resp.).} 
\label{fig:TEBDcurrentQMC}
\end{figure}

The current-voltage characteristics of the quantum dot for interaction strengths of $U/\Delta=0,4,8,12 \mbox{ and } 20$ are shown in \fig{fig:TEBDcurrent}. We plot data as obtained from QTs I and III (other QTs would give the same results but with larger error bars, as discussed in \se~\ref{ssec:extracting}). Results for the non interacting case agree with analytic results for an infinite system.~\cite{footnote3} In some regions only a likely upper bound for the steady state current can be provided. This region does not lie on the extreme end of the parameter space. It shows non trivial dependence on $U$ and $V_B$, which is discussed in detail in \se~\ref{ssec:timePerformance}. The current-voltage characteristics has an approximately semi-circular shape, with decreasing maximum as a function of interaction strength $U$. At small bias $V_B$, the current is linear in $V_B$ and agrees with the linear response result $j_{\text{lin}}=2G_0V_B$ (see also \fig{fig:TEBDcurrentQMC}). At higher bias, it departs from 
the linear response result. With increasing $U$, this departure occurs already at smaller bias $V_B$, which can be attributed to an exponential thinning of the Kondo resonance with increasing $U$.

In intermediate bias regions, we observe a flattening in the current-voltage curve. The maximum steady state current is obtained in a bias regime from $V_B\approx15\,\Delta$ to $V_B\approx19\,\Delta$. Increasing the interaction from $U=0$ to $U=12\,\Delta$ appears to shift the position of the maximum to higher bias voltages. For larger values of $U$ our data is not significant to conclude on the behavior of the position of the maximum. We find the maximum current to decrease quadratically with increasing interaction strength: $\frac{j_{max}}{\Delta}\simeq 1.675-0.003(\frac{U}{\Delta})^2$. Note that these features will likely depend on the actual reservoir DOS.

The decrease of the steady state current for high bias-voltages can be attributed to the diminishing overlap of the DOS of the two reservoirs.~\cite{PhysRevB.85.235141} Both have a semi-circular DOS with a bandwidth of $D=40\,\Delta$. In the wide-band limit, the curves behave similarly inside the low bias regime but should saturate as a function of $V_B$ for higher bias voltages (see \fig{fig:ssCurrU0}).

We discuss three simple limits. The TEBD results for the current respect the linear response ($j_{\text{lin}}$) for very low bias voltages which gives the conductance quantum $G_0$. Furthermore they respect the high bias voltage band cutoff where the current has to go to zero (here at $V_B=40\,\Delta$) due to diminishing overlap of the DOS of the reservoirs. The third limit is the non interacting case (non trivial for the used numerical method), where we obtain perfect agreement with analytical results for the thermodynamic limit.

\subsection{Comparison to previous results}\label{ssec:compare}
In the low bias region, results from other techniques are available for
the SIAM out of equilibrium. In the following we discuss our results
for various values of interaction strength $U$ together with data
previously obtained (see \fig{fig:TEBDcurrentQMC}) by
diagrammatic QMC,~\cite{PhysRevB.81.035108} fourth order Keldysh
perturbation theory,~\cite{PhysRevB.68.155310} time-dependent
DMRG,~\cite{PhysRevB.79.235336} TEBD for temperature $T=0$ (this work),
non equilibrium FRG,~\cite{jakobs_nonequilibrium_2010} non equilibrium
Cluster Perturbation Theory,~\cite{PhysRevB.86.245119} the
non equilibrium Variational Cluster Approach,~\cite{PhysRevB.86.245119,kn.li.11}
imaginary time QMC,~\cite{Dirks2012b} iterative summation of real-time
path integrals~\cite{PhysRevB.77.195316} and the linear response
result for the Kondo regime $j_{\text{lin}}=2 G_0 V_B$. All methods
work at or close to zero temperature. Some of the methods use a
wide-band limit and others a semi-circular reservoir DOS which (for equal $\Delta$) become comparable in the shown low bias
region (see \fig{fig:ssCurrU0} for a comparison). The newly obtained
TEBD results agree very well with the unbiased
dQMC~\cite{PhysRevB.81.035108} and quasi-exact
tDMRG~\cite{PhysRevB.79.235336} data. An earlier comparison including more
details but fewer techniques is available in
\tcite{1367-2630-12-4-043042}.

\subsection{Comparison to a non interacting device: Identifying correlation effects from the steady state charge current}\label{ssec:cmpU0}
\begin{figure}
  \centering
  \includegraphics[width=0.49\textwidth]{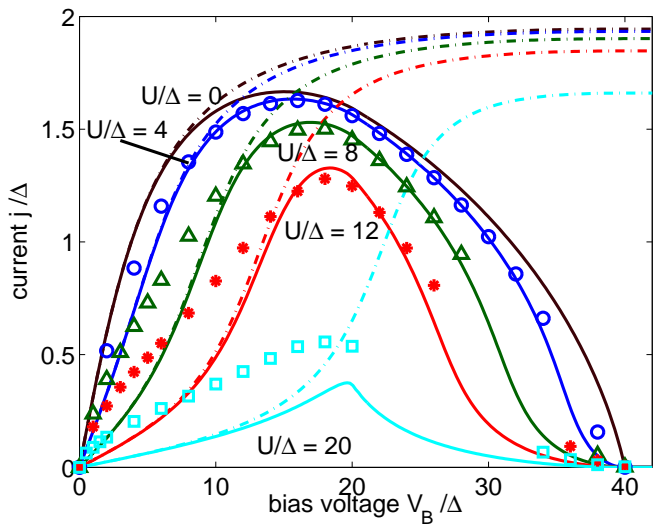}
  \caption{(Color online) Comparison of the current-voltage characteristics of a {\em non interacting}, resonant level device with on-site potential $\epsilon_f = -\frac{U}{2}$ (solid lines) with the TEBD data for the interacting quantum dot (symbols). Both devices have the same specifications with only the interaction $U$ missing in the first case. The comparison is done for four values of interaction strengths resp. on-site potentials: $\frac{U}{\Delta}=\{4,8,12,20\}$ resp. $\frac{\epsilon_f}{\Delta}=\{-2,-4,-6,-10\}$ (blue/circles, green/triangles, red/stars, cyan/squares respectively). In addition we show the $U=0$ result (black/no symbols). The dash-dotted lines indicate data for a non interacting device in the wide-band limit.} 
  \label{fig:ssCurrU0}
\end{figure}

To gain further understanding of the role of correlations, we compare the steady state current of the interacting quantum dot ($U$, on-site potential $\epsilon_f=-U/2$) to the one of a corresponding non interacting (resonant level) device with $U=0$ and on-site potential $\epsilon_f = -\frac{U}{2}$ (see \fig{fig:ssCurrU0}). Data for the resonant level device are obtained analytically.~\cite{footnote3}

From the plots in \fig{fig:ssCurrU0}, one can see clear differences in the low bias region between the non interacting and interacting device for all interaction strengths, which can be attributed to the presence of the low energy Kondo resonance in the interacting case. For low bias, the Kondo resonance fixes the linear response current to a $U$ independent constant and causes a higher current than for a non interacting quantum dot at the same on-site potential. Furthermore the curvature of the current-voltage characteristics in the low bias region is negative in the interacting case as compared to positive in the non interacting system. For larger values of $U=12\,\Delta \mbox{ and }20\,\Delta$, the negative curvature turns into a positive one in the low bias region.

For low values of interaction strength (see data for $U=4\,\Delta$) we observe deviations in both the low and high bias region. For the latter, this hints at possible many body effects which may also be important in the high bias regime. Data in the medium bias region are almost indistinguishable from the non interacting case. For high values of interaction strength the picture changes and many body effects are present in the whole bias regime.

Summing up, we find that effects of interaction are most pronounced in the low and also in the high bias regime, where a larger current is obtained than in the non interacting device. Because of the small remaining overlap of the DOS of the leads this larger current may be due to some low energy spectral weight in the interacting device, consistent with low energy excitations observed in \tcite{PhysRevB.86.245119} using a non equilibrium Variational Cluster Approach calculation.

\section{Discussion of time scales}\label{ssec:timePerformance}

In the following, we argue that Kondo correlations do not influence the steady state charge current in the parameter regime under study (large bias $V_B$ compared to Kondo scale).
However, our simulations show that depending on bias voltage and interaction strength, the steady state charge current cannot always be reached within the simulation time $\tau_E$ (see \se~\ref{ssec:extracting}), due to i) weak spots of the method (i.e. small $\tau_E$) and/or ii) long physical relaxation times. To obtain insight into physical mechanisms as well as the parameter dependence of the performance of TEBD, also relevant for future studies, it is desirable to disentangle these two effects. We identify parameter regimes with such long physical time scales to be at $U+V_B > D$ (low charge-current regime), where we find our method to perform well, as opposed to parameter regimes with high currents, where only smaller times $\tau_E$ can be reached, as shown in \app~\ref{ssec:entanglement}.

\subsection{Finite simulation size/time and Kondo correlations}\label{ssec:kondo}

At the particle-hole symmetric point of the SIAM, Kondo correlations are especially pronounced. In {\em equilibrium} they introduce a characteristic energy scale, the Kondo temperature~\cite{hewson_kondo_1997} $T_K$ which translates into a length scale of the Kondo singlet $\xi_K$, given by Bethe Ansatz~\cite{arXiv:0911.2209}
\begin{align}
\xi_K \propto \frac{v_F}{k_B T_K} \propto 2t \sqrt{\frac{2}{\Delta U}}\,e^{\frac{\pi}{8 \Delta} U}\;\mbox{.}
\label{eq:Tk}
\end{align}
Due to the exponential dependence on interaction strength, these spin correlations can not fully develop on a finite size system,~\cite{PhysRevLett.97.136604, PhysRevB.80.205114} already for moderate interaction strength. For the parameters used in this work the equilibrium Kondo correlations have a spatial extent (screening cloud) of approximately $\xi_K\approx50 \mbox{ sites for } U=4\,\Delta \mbox{, } \xi_K\approx200 \mbox{ sites for } U=8\,\Delta\mbox{, } \xi_K\approx900 \mbox{ sites for } U=12\,\Delta \mbox{ and } \xi_K\approx16000 \mbox{ sites for } U=20\,\Delta$ (see \eq{eq:Tk}). These amount to equilibrium Kondo temperatures of $T_K\approx3\cdot10^{-1}\,\Delta, 9\cdot10^{-2}\,\Delta, 2\cdot10^{-2}\,\Delta \mbox{ and } 1\cdot10^{-3}\,\Delta$ respectively.

For very small bias voltages $V_B\ll T_K$ (and large $U$), the Kondo effect introduces a large timescale. In this work however we focus on parameters for which $V_B\gg T_K$. (An exception is $U=4\,\Delta$ and $V_B<10\,\Delta$, where the Kondo cloud does fit into our finite size system.~\cite{arXiv:1211.6307}) For the parameter regime under study, recent numeric~\cite{PhysRevB.83.075107} and analytic~\cite{PhysRevLett.104.106801,PhysRevLett.108.260601,PhysRevLett.87.156802} studies provide strong indications for suppression of the equilibrium Kondo effect.

It is argued in literature that one expects a splitting of the Kondo resonance, possibly a pinning at the lead potentials~\cite{PhysRevLett.89.156801,PhysRevLett.95.126603} and/or a suppression~\cite{PhysRevB.83.075107} of the Kondo effect similar to the effect of temperature~\cite{PhysRevB.83.075107, PhysRevLett.108.260601,PhysRevB.85.201301} or magnetic field.~\cite{PhysRevB.84.245316} Renormalization group studies concluded that bias voltage is a relevant energy scale in the problem.~\cite{PhysRevLett.104.106801,PhysRevLett.108.260601,PhysRevLett.87.156802} Recent results for the electron dynamics in the steady state indicate a splitting of the Kondo resonance away from zero with bias voltage which further supports our observation that the Kondo induced timescale is not relevant for charge transport at large bias voltages.~\cite{PhysRevB.86.245119} Note that even in the presence of Kondo correlations, charge relaxation should be orders of magnitudes faster than spin relaxation.~\cite{Cohen2012}

From our current simulation we made the observation that an initial system with Kondo correlations (to be precise: their finite size remnants) as in QT II, yields the {\em same} steady state charge current (after a short, and different transient regime) as an initial system without them as in QT I. This indicates that in QT II the Kondo correlations are washed away by bias voltage. We thus conclude that although finite size systems are not able to capture the full equilibrium Kondo singlet,~\cite{PhysRevLett.97.136604} the steady state transport in the charge channel is not noticeably affected in the parameter regime under investigation. 

\subsection{Time scales in the high bias regime}\label{ssec:timescales}
\begin{figure}
\includegraphics[width=0.49\textwidth]{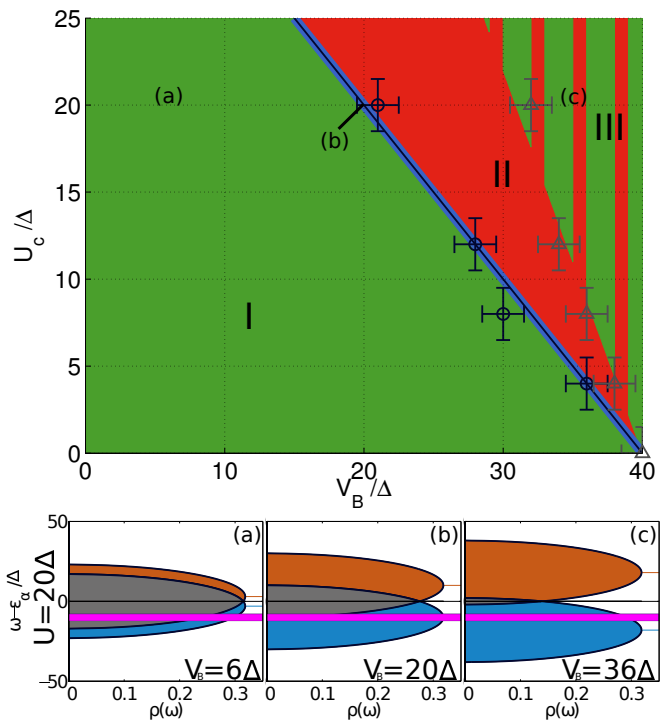}
\caption{(Color online) (top) Parameter regions in the $U-V_B$ of short (I) and long (II) physical relaxation scales as well as a regime of more complex behavior III. Data from the TEBD calculation is indicated with black and grey markers. For region II, pedestals are shown in \fig{fig:TEBDcurrent}. (bottom) Single particle DOS and single particle dot level in a Hubbard-I type picture at $U=20\,\Delta$, for (a) $V_B=6\,\Delta$, (b) $V_B=20\,\Delta$ and (c) $V_B=36\,\Delta$. The electronic DOS of the left (right) lead is shown in red (blue) and their overlap in brown. The single particle level of the quantum dot is indicated at $-\frac{U}{2}$ in magenta.}
\label{fig:timescales}
\end{figure}

We find that relaxation times in the model under discussion are strongly parameter ($U$, $V_B$) dependent. These relaxation time scales are estimated by the slope of a linear fit to the plateau region $[\tau_S,\tau_E]$ (see \se~\ref{ssec:extracting}). In particular, we identify three regions (see \fig{fig:timescales} (top)): region I is characterized by short physical relaxation times and region II exhibits longer relaxation times. Region II overlaps with the regime in which TEBD restricts us to small final simulation times $\tau_E$ (high steady state current regime, see \app~\ref{ssec:entanglement} for discussion). In region II we did not obtain a converged steady state current. In region III, the current is small and the maximum reachable simulation time (see \app~\ref{ssec:entanglement}) was large enough to determine the steady state current.

We proceed by providing an intuitive single-particle picture of the transition from region I to II in a Hubbard I type description (\fig{fig:timescales} (bottom)). Then the leads (assuming infinite reservoirs) are described by semi-circular bands of bandwidth $D$, asymmetrically shifted against each other with increasing bias voltage $V_B$. The quantum dot consists of a single (non interacting) level, located at the single particle energy $-\frac{U}{2}$. We find that the transition occurs when this single particle level of the quantum dot leaves the overlap region of both lead DOS (blue line, $U_{\text{trans}} \approx D-V_B$). We conclude that the existence of an appreciable spectral weight {\it in the overlap region} of the lead DOS leads to faster relaxation.

\section{Role of high energy states}\label{ssec:resultsDamped}
To study the role of high energy states during the time evolution we add a damping term to the propagator
\begin{align}
\hat{\mathcal{U}}(\tau) &= e^{-i\hat{\mathcal{H}}\tau(1-i\Gamma)}\,\mbox{,}
\end{align}
which gradually reduces the contribution of high energy states.

\begin{figure}
\includegraphics[width=0.49\textwidth]{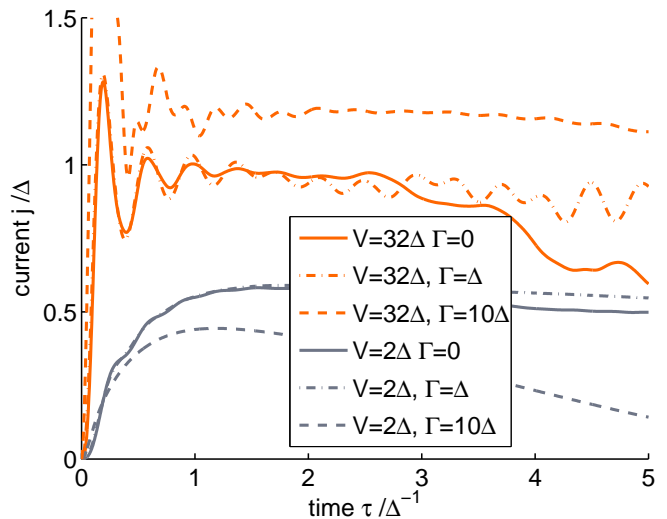}
\caption{(Color online) Effects of damping $\Gamma$ of high energy modes on the time evolution of the current ($U=0, \chi_{\text{TEBD}}=500$, QT I). Data shown are obtained for very low bias voltage ($V_B=2\,\Delta$, group of grey curves in lower part of figure) and high bias voltage ($V_B=32\,\Delta$, group of orange curves in upper part of figure). For each bias voltage we compare data obtained by a standard ($\Gamma = 0$) time evolution (full lines), data using an (empirically) optimally damped time evolution ($\Gamma=\Delta$, dash-dotted) as well as for an over-damped evolution ($\Gamma=10\,\Delta$, dashed).} 
\label{fig:dampedTimeEvolution}
\end{figure}

In \fig{fig:dampedTimeEvolution}, the effects of damping of high energy modes on the current are visualized. We show results for very low bias voltage ($V_B=2\,\Delta$) as well as high bias voltage ($V_B=32\,\Delta$). The different influence of over-damping (dashed lines) on low bias setups in contrast to high bias setups
yields insight into the role of high energy states in the two respective cases. In low bias settings, strong over-damping (here $\Gamma = 10\,\Delta$) leads to lower current while in the high bias case it leads to higher current with respect to the true one. This indicates a qualitatively different role of high energy states for these two settings.

This result can be made plausible by a simple argument. In the case of small bias voltage ($V_B\ll t$), the dominant energies should be the kinetic ones and neglecting high energy states amounts to eliminating those with highest kinetic energy. Such states contribute much to the current and neglecting them leads to a lower total current. On the other hand for very high bias voltage ($V_B>>t$), potential energy is expected to dominate. High energy states are then those with a lot of particles in the high bias reservoir. Eliminating them reduces the available state space for hopping of particles back to the side of high potential. Therefore the current is increased due to less back flow.
From a technical point of view, such an approach may reduce entanglement growth (the limiting quantity in real time evolution using matrix product states), thus reducing the required matrix dimensions of the MPS. Using such an  Ansatz however suffers from two drawbacks. i) On the one hand, we have just seen that high energy states can be important for the steady state current, and on the other hand, estimating a priori a suitable magnitude of the damping $\Gamma$ is not straightforward, since it should in principle be dynamically adjusted during the time evolution taking into account energies and truncated weight. Due to these reasons we refrain from using such an approach in general. However we show in \fig{fig:dampedTimeEvolution} that by choosing a phenomenologically good value for the damping ($\Gamma = \Delta$), one can indeed somewhat prolong the stable time evolution.

\section{Conclusions}\label{sec:conclusion}
We studied the single-impurity Anderson model out of equilibrium beyond the linear response regime by means of Density Matrix Renormalization Group. Real time evolution was performed making use of the Time Evolving Block Decimation algorithm which allows to access relevant time scales to reach the steady state. Within this framework we investigated three different quenches: i) quenching the hybridization with already applied bias voltage, ii) quenching the bias voltage and iii) quenching the hybridization at one side only.

Calculated current-voltage characteristics agree very well with established results which are available in the low bias region. We find that the period of characteristic oscillations in the time evolution of the charge current is already very well described by renormalization group results for a different model, the interacting resonant level model of spinless fermions. After an initial transient regime, where on the order of one particle is transferred through the quantum dot, the steady state current agrees among the three quenches investigated. For the identification of steady state plateaus in time dependent quantities the type of quench is however very important. We show that quenching the lead-dot tunnelings is the most suitable one, contrary to expectations whereas quenching the bias voltage results in large initial oscillations of the current. We furthermore show that limitations of the method like its inherent finite size do not pose a problem for simulations of the setup discussed here within 
accessible times. Our findings indicate that the steady state charge current is not influenced by finite size effects, hinting that incompletely developed Kondo correlations in the spin channel do not influence charge transport noticeably. We find that a large entanglement entropy correlates positively with a large steady state current amplitude. By studying a damped time evolution we find that high energy states have very different significance in the low and high bias regime respectively.

Besides reproducing the universal low bias physics we open up new perspectives for devices in which a large bias voltage is combined with a finite electronic DOS of the leads, like nano-tubes. For such devices we predict that effects of electron-electron interactions are important even at high bias voltages.

Interesting extensions within the presented approach may be the application of a gate voltage to study stability diagrams, evaluation of spin correlations which could hint on Kondo correlations, to study effects of asymmetric couplings, the interplay of bias and magnetic fields as well as to investigate correlated leads~\cite{PhysRevLett.101.236801}. On the technical side it would be interesting to evaluate whether more gently ramped quenches over a finite time interval further decrease oscillations or even entanglement and further improve the extraction of steady state data.

\begin{acknowledgments}
We gratefully acknowledge fruitful discussions with Sabine Andergassen and Steven R. White. We thank Fabian Heidrich-Meisner, Philipp Werner and Andreas Dirks for providing their data in \fig{fig:TEBDcurrentQMC}. This work was partly supported by the Austrian Science Fund (FWF) P24081-N16, and ViCoM projects F04103 and F04104. HGE thanks the KITP for hospitality. This research was supported in part by the NSF under grant No. NSF PHY05-51164. Most of the numerical calculations have been conducted at the Vienna Scientific Cluster (VSC-I$\&$II). 
\end{acknowledgments}

\appendix
\section{Method setup, convergence analysis and preliminary considerations}\label{app:preliminaryConsiderations}
\begin{figure*}
\begin{center}
  \includegraphics[width=0.99\textwidth]{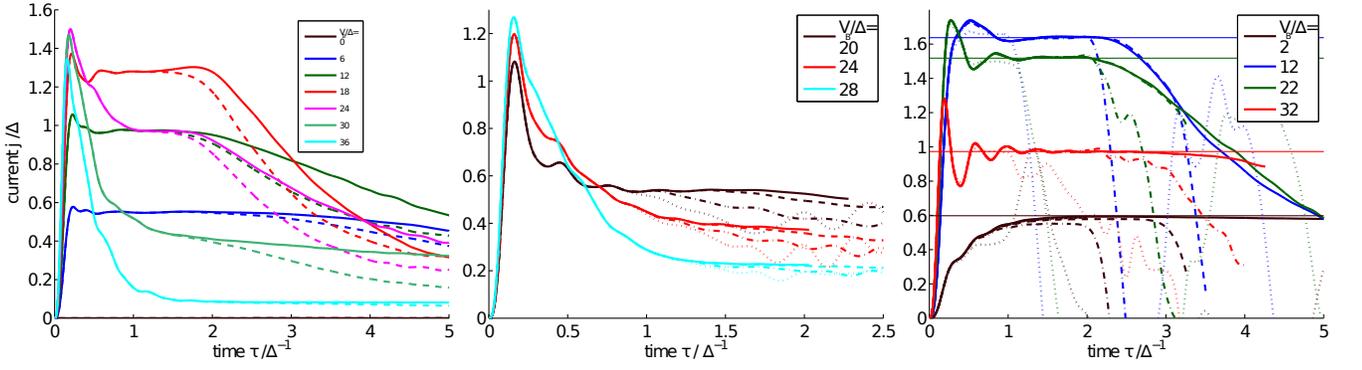}
\caption{(Color online) Convergence of the current with respect to several auxiliary numerical parameters. Left: Solid lines denote results obtained evaluating the expectation value of the current operator \eq{eq:currentOperator}, while dashed lines indicate data obtained by evaluating the time derivative of the expectation value of the particle number \eq{eq:currentDerivative} ($U=12\,\Delta$, $L=150$, $\chi_{\text{TEBD}}=2000$, QT I). Center: Matrix sizes $\chi_{\text{TEBD}}=250 \mbox{ (dotted)}, 500 \mbox{ (dash-dotted)}, 2000 \mbox{ (dashed) and } 4000 \mbox{ (solid) }$ are presented ($U = 20\,\Delta$, $L=150$, QT I). Right: We show system sizes $L=20, 40, 60, 80, 100, 120 \mbox{ and }150$ (dotted, dash-dotted, dashed, dash-dash-dot-dotted, long-dash-short-dashed, dash-gap-dashed and solid) at $U=0$, $L=150$, $\chi_{\text{TEBD}}=2000$ for QT I. The constant solid lines indicate the exact steady state currents of the respective thermodynamic system.} 
\label{fig:prelim}
\end{center}
\end{figure*}

Here we present some preliminary considerations concerning the convergence and quality of our data. Uncertainties arise from the approximations made within the method and from numerical precision.

In addition, our setup contains leads of finite size, with two effects in principle. First, this finite size affects the ground state at time zero. We will show below that the effect on the current is negligible; it converges already at much smaller lead size than used here.

Secondly, the finiteness of the reservoirs means that no energy or particle dissipation occurs and eventually the system will show oscillatory behavior. We note in passing that during our simulation time only approximately one particle traverses the quantum dot. The earliest time at which the current can be affected by the finite system size arises from a perturbation which propagates after the quench to the end of a lead and back to the quantum dot. The velocity of this signal is limited by the Lieb Robinson bound,~\cite{LiebRobinson} up to exponentially suppressed parts, and in our case is  $v\approx2t$, which can also be clearly seen in the time evolution of local charge expectation values.~\cite{NussGanahlEtcInPreparation} The perturbation will hit the left and right end of the chain and return back to the quantum dot after a time of about $\tau \approx 2 (L/2) / (2t) = L/(2t)$, i.e.\ $\tau/\Delta^{-1} \approx L \Delta/(2t) \approx 7.5 $ for $L=150$. This is far beyond the times $\tau_E$ (see Sec.\ IV A) 
up to which we calculate the steady state current, which is therefore not affected. This conclusion is confirmed by the convergence of the current with respect to system size $L$, discussed below. The measured current may however be affected by other possible errors within our approach: i) the procedure to measure it, ii) the Trotter error, and iii) the limited matrix dimension $\chi_{\text{TEBD}}$ (i.e. truncated weight).

In the following we will show that the major uncertainty arises from the limited matrix dimension $\chi_{\text{TEBD}}$, while other source are negligible. The definition of the time intervals from which the steady state current is evaluated (Sec.\ IV A) is also relevant. A similar conclusion has been drawn before in the framework of adaptive tDMRG~\cite{PhysRevB.82.205110} and for different systems in the framework of TEBD.~\cite{PhysRevB.85.235141}

\subsection{Obtaining the current}
\begin{figure}
\begin{center}
  \includegraphics[width=0.49\textwidth]{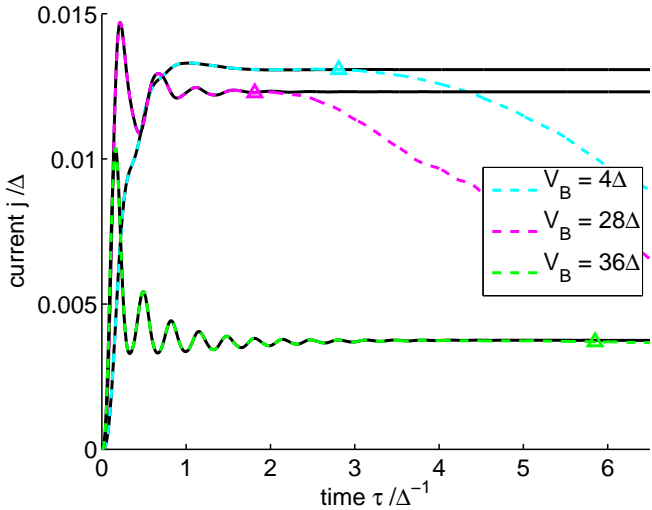}
\caption{(Color online) Exact results for the non interacting system. Comparison of the TEBD current (dashed lines) to an exact time evolution (solid black lines) for $U = 0\,\Delta$, $L=150, \chi_{\text{TEBD}}=2000$, QT I. We show results for $V_B = 4\,\Delta$ (cyan), $V_B = 28\,\Delta$ (magenta) and $V_B = 36\,\Delta$ (green). The respective maximum reliable simulation times (see \se~\ref{ssec:extracting} for definition) are indicated as triangles.} 
\label{fig:exactTimeEvolve}
\end{center}
\end{figure}

Within the TEBD time evolution the steady state current may be obtained via the expectation value of the current operator at each time step~\cite{PhysRevB.73.195304, PhysRevB.78.195317}
\begin{align*}
\hat{j}_{ij}(\tau) &= i\,t_{ij}\sum_\sigma\left(a_{i\sigma}^\dag \, a_{j\sigma} - a_{i\sigma} \, a_{j\sigma}^\dag\right)\;\mbox{,}
\end{align*}
where $i$ and $j$ denote adjacent sites and $a_{i\sigma}$ and $a_{i\sigma}^\dag$ are annihilation and creation operators for fermions on-site $i$ with spin $\sigma$ which depend at time $\tau$ and $t_{ij}$ is taken to be real. To obtain the current though the quantum dot, a symmetrized version of the inflow and outflow is used
\begin{align}
\nonumber\hat{j}(\tau) &= \frac{\hat{j}_{Lf} + \hat{j}_{fR}}{2}\\
\nonumber &=i\,\pi\,t'\sum_\sigma\Bigg(\left(f_{\sigma}^{\dag} \, c_{\text{end}\sigma}^{L} - c_{\text{end}\sigma}^{L\dag} \, f_{\sigma}\right)\\
&-\left(f_{\sigma}^{\dag} \, c_{\text{0}\sigma}^{R} - c_{\text{0}\sigma}^{R\dag} \, f_{\sigma}\right)\Bigg)\;\mbox{,}
\label{eq:currentOperator}
\end{align}
where $c_{\text{end}\sigma}^{L},c_{\text{end}\sigma}^{\dag L}$ denote operators on the last site of the left reservoir (number $74$ in \fig{fig:QD_TEBD_QT}) and $c_{\text{0}\sigma}^{R},c_{\text{0}\sigma}^{R\dag}$ denote operators on the first site of the right reservoir (number $76$ in \fig{fig:QD_TEBD_QT}).

Another way of computing the current is by calculating the time derivative of the total particle number to the left of the site under consideration
\begin{align*}
j_{i i+1}(\tau) &= \frac{d}{d\tau}\left<\sum\limits_{m=1}^{i}\sum\limits_\sigma \hat{n}_{m\sigma}(\tau)\right>\;\mbox{.}
\end{align*}
Again a symmetric combination of the dot's in- and outgoing current yields the current under consideration
\begin{align}
\nonumber j(\tau) &= \frac{1}{2}\Bigg(\frac{d}{d\tau}\left<\sum\limits_{m \in L}\sum\limits_\sigma \hat{n}_{m\sigma}(\tau)\right> \\
&+ \frac{d}{d\tau}\left<\sum\limits_{m \in L \cup f}\sum\limits_\sigma \hat{n}_{m\sigma}(\tau)\right>\Bigg)\;\mbox{.}
\label{eq:currentDerivative}
\end{align}
The current through the dot may be evaluated at each TEBD time step using \eq{eq:currentOperator} or by computing a finite difference approximation to the differential \eq{eq:currentDerivative} every two successive time steps.

Besides the expected additional source of error by evaluating the time derivative numerically, this method is expected to perform less well due to the influence of all sites in the system on the result for the current, the occupation number of each site having its own limited accuracy. A comparison of the current evaluated by means of \eq{eq:currentOperator} and \eq{eq:currentDerivative} for various values of interaction strength $U$ and applied bias voltage $V_B$ as well as all QTs (I, II, III) shows good agreement in the beginning of the time evolution (see \fig{fig:prelim} (left)). Due to an accumulation of errors in the particle number expectation values of the individual sites, the results start to deviate at some time $\tau^{(2)}_E$. We do not use results beyond $\tau^{(2)}_E$ (see the discussion in \se~\ref{ssec:extracting}. Numerical values of all steady state currents will be obtained using the current operator (\eq{eq:currentOperator}) which yields a much more stable estimator.

\subsection{Finite size effects: $L$}
In this section we discuss the dependence of the results for the current on the length of the system $L$.~\cite{ANDP:ANDP201000017,PhysRevB.85.235141} We quench both dot-lead tunnelings (i.e. QT I). The qualitative behavior for the other QTs (II and III) is virtually identical. Results for the steady state current for system sizes of $L=20,40,60,80,100,120 \mbox{ and } 150$ sites are shown in \fig{fig:prelim} (right) in the non interacting case. We find that the final results for the steady state current agree with the analytically available results for an infinite system in all cases within the numerical error. This ensures a reliable determination of steady state properties even on finite size systems. As mentioned before, the system size limits the maximum simulation time due to signals back-propagating from the borders. In the main part of this work all calculations are performed for a system size of $L=150$ to provide a nice long plateau (maximum simulation time) in the steady state current. It has been 
noted in \tcite{arXiv:0601389} that in the particle hole symmetric half filled model the steady state current is independent of system size. A detailed discussion of finite size and time scales in a model of spinless fermions can be found in \tcite{ANDP:ANDP201000017}.

For completeness we note that it is possible to extend the available simulation time, when it is limited by the hard boundary conditions of the leads, by applying modified boundary conditions.~\cite{PhysRevLett.71.4283,0295-5075-73-2-246,PhysRevB.78.195317,PhysRevLett.101.236801} Exponentially decreasing the matrix elements of the Hamiltonian towards the end of the reservoirs ultimately corresponds to a Wilson chain with logarithmic discretization.~\cite{0295-5075-73-2-246} In this work we do not apply any modified boundary conditions because our simulation time is not limited by the size of the chains but the TEBD matrix dimension $\chi_{\text{TEBD}}$.

\subsection{Trotter error: $\delta\tau$}
The Trotter error grows only linearly with simulation time,~\cite{RevModPhys.77.259, PhysRevB.82.205110} and can be controlled by choosing sufficiently small $\delta\tau$. Therefore usually the contribution to the total error arising due to the Trotter approximation is negligible with respect to other approximations. We investigated the influence of the Trotter decomposition on the current. Results for $\delta\tau/t^{-1}=\{0.01, 0.05, 0.1\}$ were found to agree to within $5\cdot10^{-5}$. We do not plot the results because they all lie on top of each other. A good value for the time step was found to be $\delta \tau /t^{-1}=0.05$ which was used in the main section of the paper.

\subsection{MPS matrix dimension: $\chi$}
The quality of the TEBD results is predominantly determined by the maximum matrix size $\chi_{\text{TEBD}}$ used. A bigger $\chi_{\text{TEBD}}$ leads to fewer discarded states (i.e. less truncated weight of the reduced density matrices) during the truncation and therefore to a systematically better approximation.~\cite{PhysRevB.70.121302} The truncated weight is defined by~\cite{RevModPhys.77.259}
\begin{align}
 \epsilon = 1-\sum\limits_{\gamma = 1}^\chi \lambda_{\gamma}^2\;\mbox{,}
 \label{eq:tw}
\end{align}
where $\lambda_{\gamma}^2$ denote the eigenvalues of the reduced density matrices. This quantity is zero if no truncation is done. The computational cost of the TEBD algorithm scales essentially like~\cite{RevModPhys.77.259}
\begin{align*}
\mbox{cost} \propto L(d\;\chi_{\text{TEBD}})^3\;\mbox{,}
\end{align*}
where $L$ is the length of the chain and $d=4$ the size of the local fermionic Hilbert space. Therefore it is essential to keep $\chi_{\text{TEBD}}$ as low as possible. During the simulations we noticed that at a certain time (long before signals propagating back from the ends of the chain would reach the quantum dot) the truncated weight starts to grow quickly and the results become unstable,~\cite{PhysRevE.71.036102} causing a decaying current. The effects of enlarging $\chi_{\text{TEBD}}$ are shown in \fig{fig:prelim} (center). As the data indicates, the effect of increasing $\chi_{\text{TEBD}}$ is to make larger simulation times accessible, before the simulation breaks down due to accumulation of truncated weight. Remarkably, no spurious quasi steady state is entered when $\chi_{\text{TEBD}}$ is relatively small. The overall shape of the current appears to be unaffected by enlarging $\chi_{\text{TEBD}}$, making reliable predictions for $\chi_{\text{TEBD}}=2000$ possible. We checked our results in all 
parameter regions for convergence. In the main part of the paper we always used $\chi_{\text{TEBD}}=2000$ as a good compromise between run time and accuracy.

\subsection{Comparison to analytical results}\label{ssec:ctar}

\begin{figure}
\includegraphics[width=0.49\textwidth]{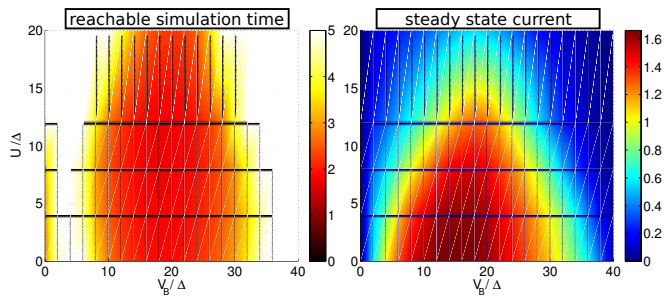}
\caption{(Color online) Maximum simulation time reachable (QT I, $\chi_{\text{TEBD}}=2000$) due to accumulation of entanglement entropy (left) and steady state current (right). The left plot shows the time until a truncated weight of $\epsilon_c=5\cdot10^{-5}$ is reached at any bond of the chain for the first time. The right figure corresponds to the data in \fig{fig:TEBDcurrent}. Note the inverted color-scale in the left image, dark regions correspond to low values of the maximum time reachable.}
\label{fig:currentTWcorrelation}
\end{figure}

In the non interacting setup $U = 0$, we compare TEBD data to results from an exact time evolution (see \fig{fig:exactTimeEvolve}). The exact time evolution was obtained for the same system parameters by time evolving the single particle density matrix. We find that the TEBD time evolution is reliable up to a system parameter dependent time. This time (triangles) again is in accordance with the criterion for the maximum reachable simulation time as defined in \se~\ref{ssec:extracting} and has a non trivial dependence on bias voltage and interaction strength.

The non interacting system is non trivial for the TEBD method. Our data for the entanglement entropy and the truncated weight at low-, medium-, as well as high bias voltages for increasing interaction strength~\cite{NussGanahlEtcInPreparation} indicate that indeed the $U=0$ case does not exhibit a particular low entanglement or truncated weight in comparison with higher interaction strength. Since we reproduce the exact analytic steady state current in the non interacting case we conclude that the agreement with exact results is not a peculiarity of the non interacting system and our way of data extraction can be applied to finite values of $U$.

\subsection{Setup}
Based on the above considerations all data in the main part of the paper were obtained for the following parameters: i) The ground state was obtained by DMRG using a matrix size $\chi_{DMRG}=400$ and undergoing $10$ sweeps of two-site DMRG before switching to $40$ runs for single-site DMRG.  ii) The model consists of $L=150$ sites. Upon performing one of the three above described quenches (I, II or III) we used bias voltages in a range of $V_B/\Delta=(0, 42)$. We always started from an overall half filled system in the canonical ensemble with total $S^z=0$ and alternating up and down spins are chosen from left to right. iii) The time evolution was performed using a TEBD matrix size of $\chi_{\text{TEBD}}=2000$, a trotter step of $\delta\tau/\Delta^ {-1}=0.005$ and evolving for $1000$ time steps which yielded a final simulation time of $T/\Delta^{-1}=5$. Requiring a maximum truncated weight of $\epsilon_c=10^{-15}$ we dynamically adjusted the size of the TEBD matrices with a maximum matrix size of $\chi_{\
text{TEBD}}$. We measured observables at each time step.

\section{Correlation of entropy and steady state current}\label{ssec:entanglement}
 
The major limiting factor for time evolution using TEBD is the increase of bipartite entanglement~\cite{Schollwoeck201196}
\begin{align*}
 S_i = -\text{tr}\left(\hat{\rho}_L\,\ln{(\hat{\rho}_L)}\right)= -\text{tr}\left(\hat{\rho}_R\,\ln{(\hat{\rho}_R)}\right)\;\mbox{,}
\end{align*}
where $\hat{\rho}_{L/R}$ denotes the reduced density matrix to the left ($L$) and to the right ($R$) of a bipartition at bond $i$.

Using a maximum matrix dimension $\chi_{\text{TEBD}}$, we stop the simulation (for \fig{fig:currentTWcorrelation}) whenever the truncated weight at any bond exceeds a threshold value of $\epsilon_c=5\cdot 10^{-5}$, which defines our maximum simulation time $\tau^{(1)}_{E}$ (see \se~\ref{ssec:extracting}). In \fig{fig:currentTWcorrelation}, we plot $\tau^{(1)}_{E}$ as a function of $U$ and $V_B$ (left) and compare it to the magnitude of the steady state current for the same parameters (right).

From our data we conclude that reachable simulation times due to accumulation of entanglement (and thus truncated weight) are non monotonic in $U$ and $V_B$ but can be characterized roughly by the magnitude of current in the system. We find this behavior to be generic to all investigated QTs.

\bibliography{nSIAM}{}

\begin{thebibliography}{77}
\bibitem[{foo({\natexlab{a}})}]{footnote1}
\bibinfo{note}{In the wide-band limit a featureless (constant) electronic
  density of states of the reservoirs is assumed which has a bandwidth much
  larger than the characteristic energy scales of the system coupled to the
  reservoir. The retarded Green's function is then given by
  $G^R(\omega)=-\frac{1}{2W} \text{ln}\left(\frac{\omega-W}{\omega+W}\right)$
  where $W$ is the half bandwidth.}
\bibitem[{foo({\natexlab{b}})}]{footnote2}
\bibinfo{note}{Which is in fact a ``quasi'' steady state due to the fact that
  the system is of finite size. We however show that it corresponds to the true
  steady state of the corresponding thermodynamic system.}
\end{thebibliography}

\end{document}